\documentclass[english,prl,twocolumn,superscriptaddress]{revtex4}
\usepackage{times}
\usepackage[T1]{fontenc}
\usepackage[latin1]{inputenc}
\usepackage{amsmath}
\usepackage{amssymb}
\usepackage{float}
\usepackage{color}
\usepackage{graphicx}
\usepackage{amssymb}
\makeatletter
\usepackage{babel}
\makeatother	

\begin{document}

\title{Cooper pair insulator in amorphous films induced by nanometer-scale thickness variations}

\date{\today{}}
\author{S. M. Hollen}
\author{H. Q. Nguyen}
\author{E. Rudisaile}
\affiliation{Department of Physics, Brown University, Providence, RI 02912}
\author{M. D. Stewart, Jr.}
\affiliation{Department of Physics, Brown University, Providence, RI 02912}
\author{J. Shainline}
\affiliation{Department of Physics, Brown University, Providence, RI 02912}
\author{J. M. Xu}
\affiliation{Department of Physics, Brown University, Providence, RI 02912}
\affiliation{Division of Engineering, Brown University, Providence, RI 02912}
\author{J. M. Valles, Jr.}
\affiliation{Department of Physics, Brown University, Providence, RI 02912}

\maketitle

\textbf{
Unusual transport properties of superconducting (SC) materials, such as the under doped cuprates\cite{Hackl:NJPhys2010}, low dimensional superconductors in strong magnetic fields\cite{Uji:PRL2006}, and insulating films near the Insulator Superconductor Transition (IST)\cite{Gantmakher:PUsp2010}, have been attributed to the formation of inhomogeneous phases.  Difficulty correlating the behaviors with observations of the inhomogeneities make these connections uncertain. Of primary interest here are proposals that insulating films near the IST, which show an activated resistance and giant positive magnetoresistance, contain islands of Cooper Pairs (CPs)\cite{Kowal:SSC1994,Dubi:Nature2007,Feigelman:PRL2007,Vinokur:Nature2008}. Here we present evidence that these types of inhomogeneities are essential to such an insulating phase in amorphous Bi (a-Bi) films deposited on substrates patterned with nanometer-sized holes.  The patterning induces film thickness variations, and corresponding coupling constant variations, that transform the composition of the insulator from localized electrons to CPs. Analyses near the thickness-tuned ISTs of films on nine different substrates show that weak links between SC islands dominate the transport.  In particular, the ISTs all occur when the link resistance approaches the resistance quantum for pairs. These observations lead to a detailed picture of CPs localized by spatial variations of the superconducting coupling constant.}

A central issue in the field of insulator to superconductor transitions concerns the nature of the insulating phase.  In most homogeneously disordered thin films, the insulator is composed of unpaired electrons that are weakly localized by disorder\cite{Haviland:PRL1989,Valles:PhysicaB1994,Finkelshtein:JETP1987,Belitz:PRB1989}.  A less familiar state which supports remnants of superconductivity appears in other homogeneously disordered films, such as In Oxide and TiN\cite{Sambandamurthy:PRL2004,Baturina:PRL2007}.  This so-called Cooper Pair Insulator (CPI) state exhibits activated transport and a giant positive magnetoresistance, which suggest that localized Cooper pairs participate in transport. How this bosonic insulating state develops has not been determined. Theoretical descriptions of the CPI phase often presume that disorder or magnetic field creates islands of Cooper Pairs (CPs) embedded in a non-SC background\cite{Falco:PRL2010,Vinokur:Nature2008,Dubi:PRB2006,Gantmakher:JETP1998}. Indeed, scanning tunneling spectroscopy experiments show evidence of incipient puddling in some films close to the IST\cite{Sacepe:PRL2008}. Since Cooper pairs are not evident in most other insulating homogeneously disordered thin films, the factors responsible for the change in behavior are generally unclear. 

\begin{figure}
\begin{center}
\includegraphics[width=\columnwidth,keepaspectratio]{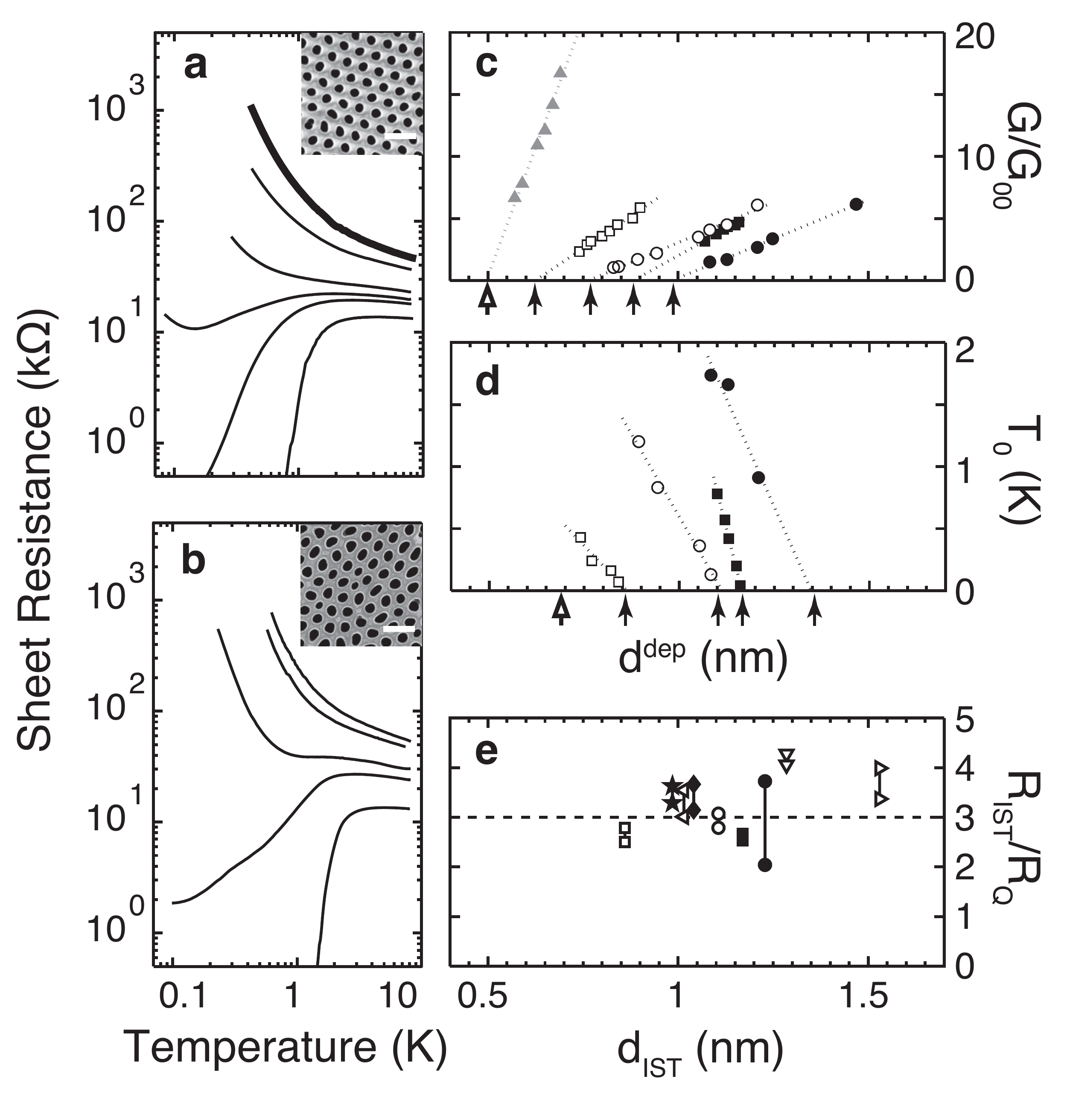}
\caption{{\bf Characteristics of thin films on nanohoneycomb substrates} (a) and (b) Insulator Superconductor Transitions tuned by thickness of a-Bi on more and less ordered hole arrays.  Insets: post-experiment SEM images. Center-to-center hole spacing: 100 $\pm$ 5 nm. Radii: 27 and 28 $\pm$ 3 nm. The scale bar spans 200 nm.  (c) Normalized conductance ($G_{00}=1/81k\Omega$) and (d) activation energies for films on four different NHC substrates.  Similar symbols represent data from experiments on side-by-side NHC substrates. Unfilled and filled symbols represent data from more and less ordered hole arrays, respectively. Data from the experiment in (a) and (b) are shown as circles. For squares, $r_{hole}=$ 16 (unfilled) and 15 (filled) $\pm$  3nm. Reference film data are shown as grey triangles in (c).  The dotted lines are least-squares linear fits to the data, and the arrows indicate critical thicknesses for conductivity and SC.  Open arrows indicate critical thicknesses on a reference film: 0.5 nm for conductivity and 0.7 nm for SC. (e) Normal state ($T=8K$) critical sheet resistance for the last insulating and first SC films of the IST of nine different NHC substrates.  The horizontal dashed line at 3$R_\mathrm{Q}$ is a guide to the eye. 
\label{cap:Gvsd}}
\end{center}
\end{figure}

Recently, it was established that amorphous Bi films can be induced to form a CPI state instead of the unpaired insulator state by depositing them on an anodized Al oxide substrate patterned with a nanohoneycomb (NHC) array of holes (see Fig. \ref{cap:Gvsd}(a) and (b) insets). The holey substrates have non-uniform topography that causes thickness and, consequently, superconducting coupling constant variations throughout the film.  The insulating NHC a-Bi films near the IST exhibit the characteristic transport signatures of the CPI phase\cite{Stewart:Science2007,Nguyen:PRL2009} .  Moreover, the insulators are confirmed to contain Cooper pairs through Little-Parks like magnetoresistance oscillations induced by the hole array\cite{Stewart:Science2007}. These films provide a unique example of a system developing a CPI phase as a result of the introduction of inhomogeneities. Establishing that well defined islands form at the IST will put important constraints on models of the Cooper Pair Insulator phase.

\begin{figure}
\begin{center}
\includegraphics[width=\columnwidth,keepaspectratio]{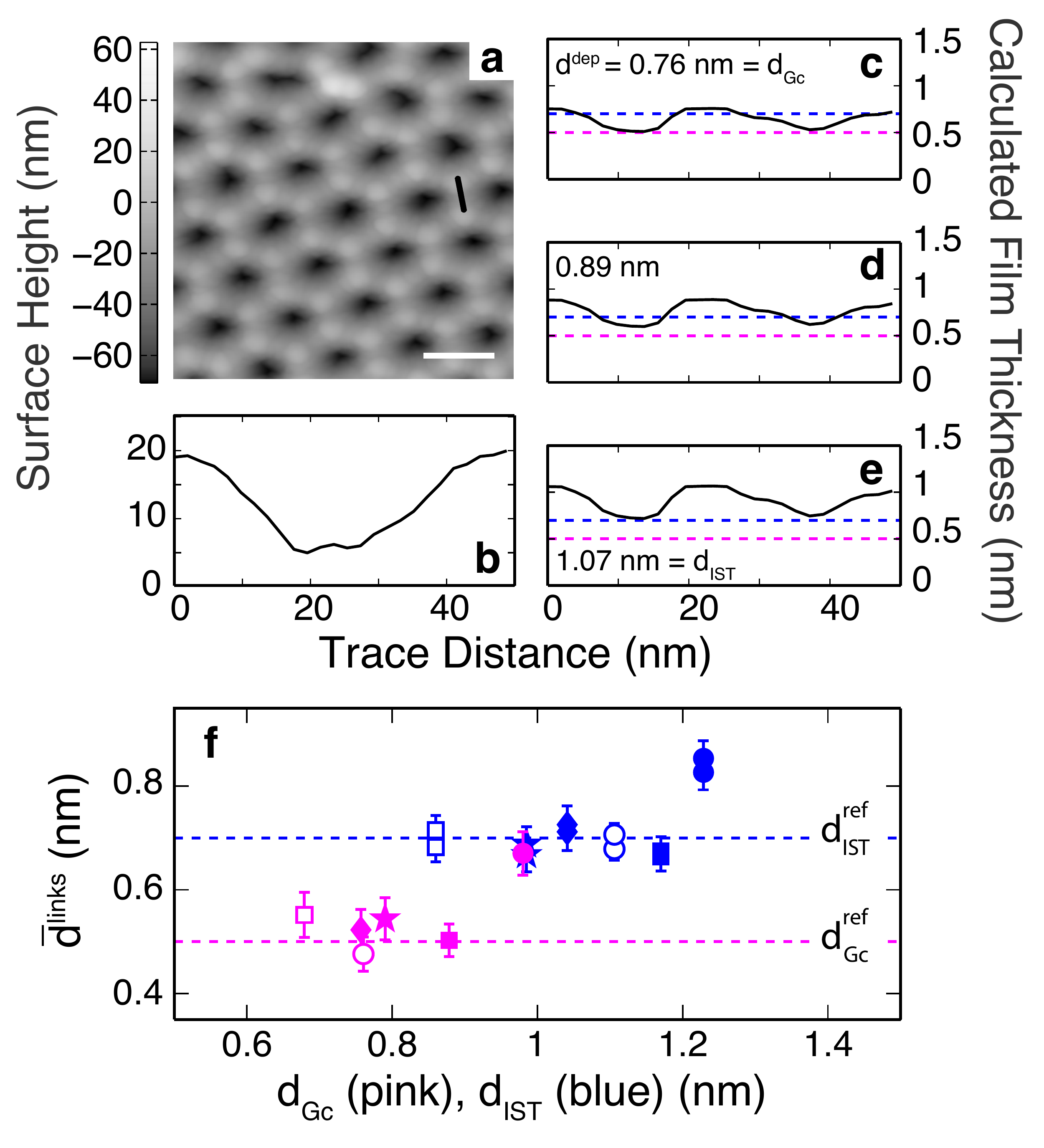}
\caption{{\bf NHC film topography} (a) AFM image showing the topography of the substrate in Fig. \ref{cap:Gvsd}(a). The scale bar (white line) spans 100 nm. (b) Height profile along the trace (black line) in (a).  (c)-(e) Film thicknesses along the trace in (a) for $d^\mathrm{dep}$ at (c) the onset of conduction, (d) a representative insulating film (bold in Fig. \ref{cap:Gvsd}(a)), and (e) the IST.  Pink and blue lines indicate the critical thicknesses for conductivity and SC, respectively, for a reference film.   (f) Average link thickness for NHC films at their transitions to conductivity ($d_\mathrm{Gc}$, pink) and SC ($d_\mathrm{IST}$, blue) for six different substrates. Critical thicknesses for SC are shown by plotting the link thicknesses of the last insulating and first SC films spanning the IST.  
\label{cap:fig2v3}}
\end{center}
\end{figure}

\begin{figure*}
\begin{center}
\includegraphics[width=2\columnwidth,keepaspectratio]{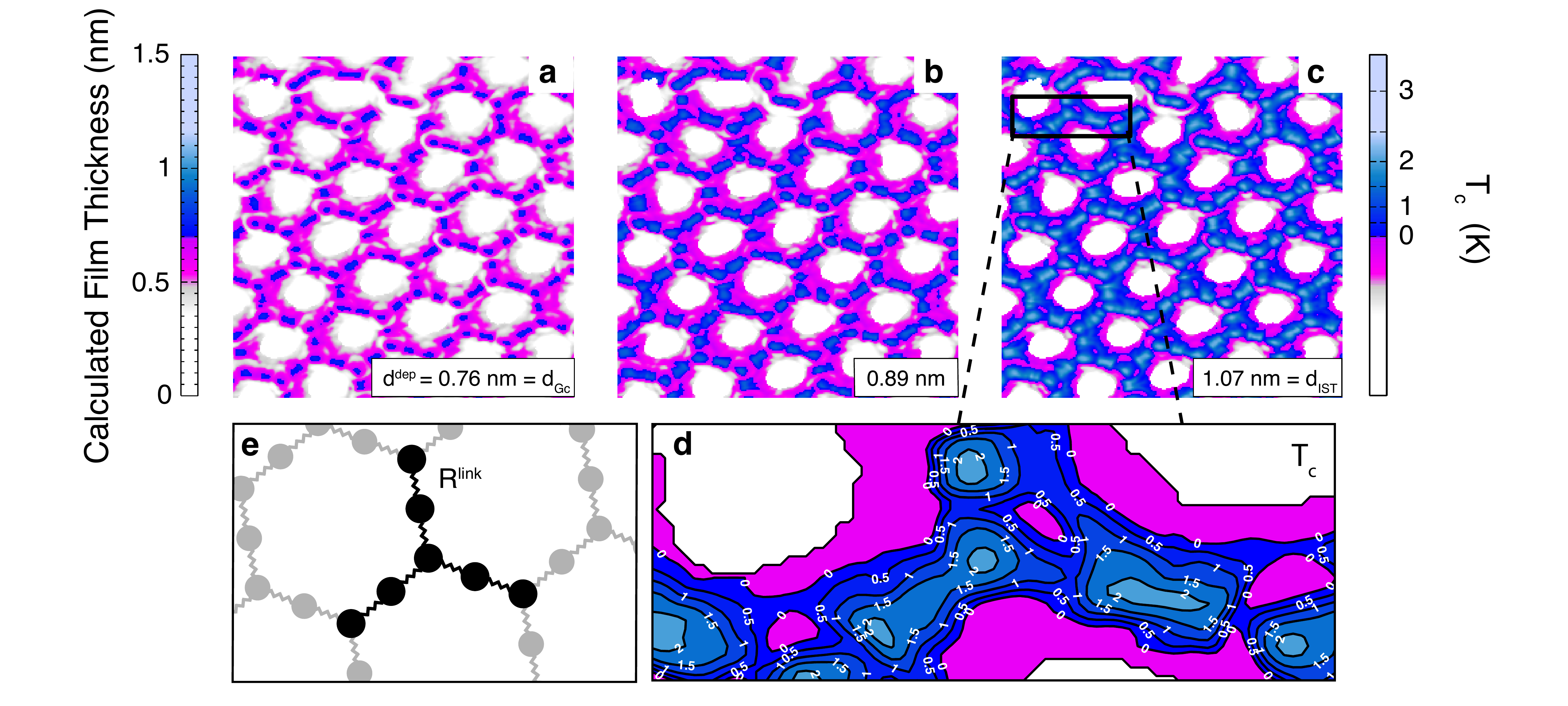}
\caption{{\bf Film thickness variations in NHC films} Calculated thickness variations in films deposited on the NHC substrate of Fig. \ref{cap:Gvsd}(a) and Fig. \ref{cap:fig2v3}(a).  Film thickness colors represent behavior on a reference film: white is non-conductive ($d<d_\mathrm{Gc}^\mathrm{ref}$), pink is conductive ($d>d_\mathrm{Gc}^\mathrm{ref}$), and blue superconduct ($d>d_\mathrm{IST}^\mathrm{ref}$).  (a)-(c) Film thickness variation plots (a) at the onset of conduction, (b) for a representative insulating film, bold in Fig. \ref{cap:Gvsd}(a), and (c) at the IST.  The critical temperatures for SC in reference films ($T_\mathrm{c}$) of these thicknesses are shown on the right. $T_\mathrm{c}=0$ marks the IST and the maximum $T_\mathrm{c}$ seen in each of the panels is (a) 0.4K, (b) 1.3K, and (c) 2.2K. (d) $T_\mathrm{c}$ (in Kelvin) contours shown for a section of the film at $d^\mathrm{dep}=d_\mathrm{IST}$. (e) Illustration of a wire array of resistors, $R^\mathrm{link}$, connecting SC islands. The black segments show the basic element of the array.
 \label{cap:dvar}}
\end{center}
\end{figure*}

Here we focus on the systematics of thickness tuned ISTs in these films. We pay particular attention to how the critical thicknesses for the onset of conduction ($d_\mathrm{Gc}$) and superconduction ($d_\mathrm{IST}$) vary with substrate qualities and also point out the relative invariance of the critical normal state sheet resistance ($R_\mathrm{IST}$).
The ISTs, shown in Fig. \ref{cap:Gvsd}(a) and (b), differ significantly from those of their unpatterned, or ``reference,'' counterparts (see Ref. \cite{Haviland:PRL1989} for an example).  The $R(T)$ evolve with deposited thickness, $d^\mathrm{dep}$, from a monotonic rise as $T\rightarrow 0$ that is exponential rather than logarithmic, to develop a reentrant dip not seen in reference films, to finally drop steadily into the SC state. The exponential rise indicates thermally activated transport, $R(T)\propto e^{T_{0}/T}$\cite{Stewart:Science2007}.  For all the ISTs, the normal state sheet conductance, $G(8 K)=1/R(8 K)$ grows linearly with deposited film thickness, $d^\mathrm{dep}$, starting at a critical thickness, $d_\mathrm{Gc}$(Fig. \ref{cap:Gvsd}(c)). This linear behavior reflects homogeneous rather than granular film growth, which would appear as an exponential dependence\cite{Dynes:PRL1978}.  Consistently, the slope, $dG(8 K)/dd^\mathrm{dep}$, is smaller and the intercept, $d_\mathrm{Gc}$, is larger for the NHC films compared to reference films. $T_0$ decreases continuously on approaching the IST (also shown  in Refs. \cite{Stewart:Science2007, Stewart:PRB2008, Nguyen:PRL2009}).  We define the critical thickness for the IST, $d_\mathrm{IST}$, as the thickness at which $T_0$ linearly extrapolates to zero.  Notice that similar to $d_\mathrm{Gc}$, $d_\mathrm{IST}$ is larger for NHC films than reference films.  Finally, $R_\mathrm{IST}$ shows little variation over nine different substrates that exhibit a wide range of $d_\mathrm{IST}$ (see Fig. \ref{cap:Gvsd}(e)). The $R_\mathrm{IST}$s cluster around $R_\mathrm{IST}\approx 3R_\mathrm{Q}$, where $R_\mathrm{Q}=h/(2e)^{2}$, the quantum of resistance for pairs, is the critical sheet resistance for SC in reference films.

Topographical analyses of AAO substrates using AFM images show that the minimum film thickness in critical NHC films (at $d^\mathrm{dep}=d_\mathrm{Gc}$ and $d_\mathrm{IST}$) correlates with critical thicknesses in reference films, $d_\mathrm{Gc}^\mathrm{ref}$ and $d_\mathrm{IST}^\mathrm{ref}$.
  As pointed out previously\cite{Stewart:PhysicaC2009}, the NHC substrate topography $h(x,y)$ shows regular variations around holes, which makes the thermally evaporated film thickness a function of position, $d(x,y)$.   For a uniform flux of atoms impinging parallel to the average normal to the substrate these are given by:
\begin{equation}\label{eq:dvar} d(x,y)=d^\mathrm{dep}\frac{1}{\sqrt{1+(\nabla h)^{2}}}.\end{equation}
Fig. \ref{cap:fig2v3}(a) shows the topography of the AAO substrate. Six peaks, visible as bright spots, surround each of the holes.  A line scan between two neighboring peaks shows a single central minimum in the height (Fig. \ref{cap:fig2v3}(b)). Using Eqn. \ref{eq:dvar}, we calculate thickness profiles along this link corresponding to the onset of conduction ($d^\mathrm{dep}=d_\mathrm{Gc}$), an insulating film ($d_\mathrm{Gc}<d^\mathrm{dep}<d_\mathrm{IST}$, bold in Fig. \ref{cap:Gvsd}(a)), and the onset of SC ($d^\mathrm{dep}=d_\mathrm{IST}$) (see Figs. \ref{cap:fig2v3}(c-e)).   All show two minima that correspond to the steepest regions of the height profile. At the critical thicknesses, $d_\mathrm{Gc}$ and $d_\mathrm{IST}$, the minima coincide with critical thicknesses in reference films, $d_\mathrm{Gc}^\mathrm{ref}$ (pink line at 0.5 nm) and $d_\mathrm{IST}^\mathrm{ref}$ (blue line at 0.7 nm).  These observations intimate that NHC films conduct or SC when the thinnest regions, the weak links, reach the appropriate critical thickness. Calculations of the weak link thicknesses averaged over each of the six substrates analyzed, $\bar{d}^\mathrm{links}$, strongly support this suggestion.  Fig. \ref{cap:fig2v3}(f) compares $\bar{d}^\mathrm{links}$ at NHC film critical thicknesses with the critical thicknesses for conductivity and SC of reference films.  It is evident that for all but one of the substrates $\bar{d}^\mathrm{links}$ at critical $d^\mathrm{dep}$ corresponds to critical thicknesses in reference films.

Full two-dimensional maps of the local film thickness, $d(x,y)$ reveal a compelling picture: SC and normal regions coexist in NHC films through the IST (Fig. 3). These maps were generated by applying Eqn. \ref{eq:dvar} to the AFM image in Fig. \ref{cap:fig2v3}(a) for the series of depositions in Fig. \ref{cap:fig2v3}(c), (d), and (e). The color scale indicates regions that are SC, resistive, and insulating according to their behavior at those thicknesses in a reference film. It is evident that SC islands form and grow where the substrate is flattest, such that there are generally 12 islands surrounding each hole, one for each peak and valley. Through the IST, the SC islands are larger than the coherence length in these films, $\xi\approx$ 15 nm. The right axis of the figure further converts $d(x,y)$ to a local $T_\mathrm{c}$ map, $T_\mathrm{c}(x,y)$, using the relationship between $T_\mathrm{c}^\mathrm{ref}$ and $d^\mathrm{ref}$ observed here and in many amorphous film systems\cite{Haviland:PRL1989, Lee:PRL1990}: $T_\mathrm{c}^\mathrm{ref}=T_\mathrm{c}^\mathrm{bulk}(1-d_\mathrm{IST}^\mathrm{ref}/d^\mathrm{dep})$. For a-Bi, $T_c^{bulk}=6$ K and $d_{IST}^{ref}=0.7$ nm\cite{Jay:thesis}. Note that the $T_\mathrm{c}$ maps are better thought of as coupling constant maps since they do not include proximity effects. Each of the islands has a larger $T_\mathrm{c}$ at its center than on its edges and resistive film regions surround the islands. With increasing thickness, the intervening resistive regions shrink until the SC islands coalesce (see Supplemental Information).  The $T_\mathrm{c}$ contour plot, Fig. \ref{cap:dvar}(d), emphasizes that $T_\mathrm{c}$ in the regions between islands at their coalescence is finite, but much less than the maximum at the islands' centers. Thus, these regions likely act as weak links. 

The color maps suggest a network model for NHC films that provides a clue to the origin of the apparently universal normal state resistance at the IST.  In the model, the SC islands become highly conductive regions connected by resistive weak links as shown in Fig. \ref{cap:dvar}(e).  The resistance of the basic building block of this network, which is equivalent to the macroscopic sheet resistance, is $3R^\mathrm{link}$.  Thus, according to Fig. \ref{cap:Gvsd}(f), at the IST $R_\mathrm{IST}^\mathrm{link} \approx R_\mathrm{Q}$.   Interestingly, theory\cite{Fazio:PhysRep2001} and experiment\cite{Takahide:PRL2000} maintain that arrays of small, resistively shunted Josephson Junctions in the extreme quantum limit undergo an IST for $R_\mathrm{N}\approx R_\mathrm{Q}$.  Applying these dissipation driven IST models to NHC films seems plausible as the SC islands are small enough to be in the extreme quantum limit and the resistive regions surrounding the islands can act as the dissipative elements. 

We have shown that the CPI phase in NHC a-Bi films is composed of superconducting islands.  These islands result from variations in the film thickness on scales larger than $\xi$ that lead to inhomogeneities in the superconducting coupling constant. Our observations support conjectures that CP islanding gives rise to the activated resistance and giant magnetoresistance observed in this and other thin film systems, such as InOx and TiN. For these structure-free films, however, an alternative mechanism for island formation is required. Whether this mechanism is spontaneous electron-electron interaction driven phase separation\cite{Galitski:PRL2005}, the fractal nature of the electronic wavefunctions close to the metal to insulator transition\cite{Feigelman:PRL2007}, a random Gaussian potential\cite{Falco:PRL2010}, underlying structural inhomogeneities in the material\cite{Ghosal:PRL1998, Dubi:Nature2007}, or yet to be determined is an open question. We anticipate that the detailed picture of the superconducting inhomogeneities presented here will aid microscopic calculations of these and other characteristics, such as the tunneling density of states, of the CPI state.

{\flushleft{\bf Methods:}}

NHC a-Bi thin films are grown by thermally evaporating Sb and then Bi onto an Anodized Aluminum Oxide (AAO)\cite{Yin:APL2001} substrate held at 8K in a dilution refrigerator cryostat and measured \emph{in situ}. Reference films are grown simultaneously on a fire-polished glass substrate.  Film thicknesses are measured with a quartz crystal microbalance. The semiconducting Sb underlayer ($d\simeq$1 nm) promotes the surface wetting of the subsequently deposited Bi.  Previous STM analyses of quench condensed metal films on semiconductors indicate that the metal atoms stick where they land to form a homogeneous amorphous film\cite{Ekinci:PRB1998}.  Thus, the films assume the holey geometry of the underlying AAO substrate (see inset Fig. \ref{cap:Gvsd}(a) and (b)).  In the nine different AAO substrates used in these experiments the average hole radii range from 14 to 28 $\pm$ 3 nm with average center-to-center hole spacing of 100 $\pm$ 5 nm in each substrate.   The order in the hole arrangement was intentionally varied using two different anodization processes (see insets Fig. \ref{cap:Gvsd}(a) vs. (b)).  Sheet resistances as a function of temperature, $R(T)$, were measured on 1.5 mm square areas defined by predeposited Ge/Au contacts using four point AC and DC methods. The geometry of each of the nine hole arrays was characterized using Scanning Electron Microscope images and the surface topography of six of the nine was measured using tapping mode AFM over a 1$\mu$m$^{2}$ area.

{\flushleft{\bf Acknowledgements:}} 

We are thankful to acknowledge helpful conversations with D. Feldman, N. Trivedi, T. Baturina, and A. Goldman. This work was supported by the NSF through Grant No. DMR-0605797 and No. DMR-0907357, by the AFRL, and by the ONR.

\clearpage
\begin{figure*}
\begin{center}
\includegraphics[width=2\columnwidth,keepaspectratio,clip=true, viewport=.75in .75in 7.75in 10.25in]{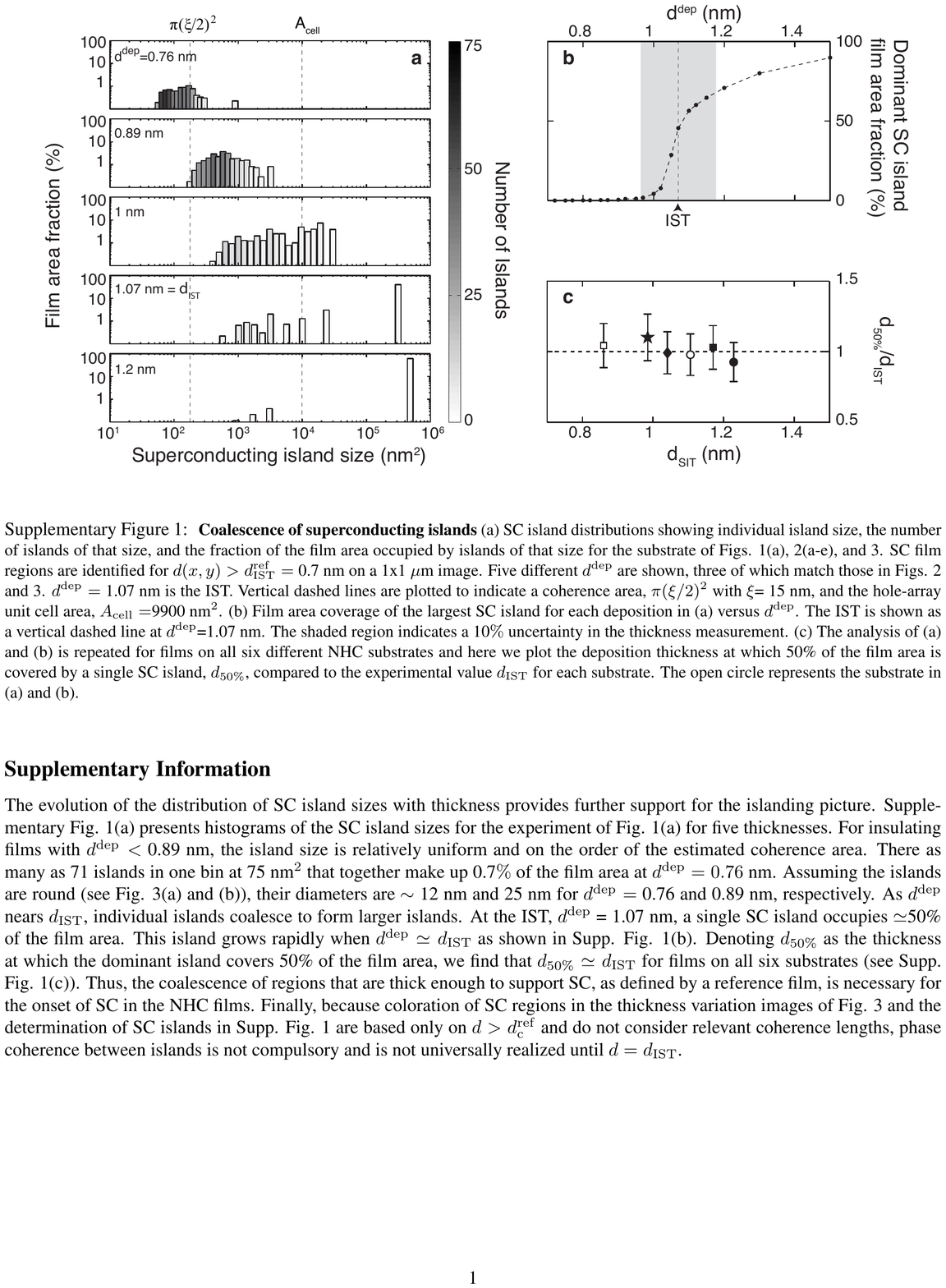}
\end{center}
\end{figure*}

\end{document}